\documentstyle[11pt,moriond,epsfig]{article}

\newcommand \beq{\begin{eqnarray}}
\newcommand \eeq{\end{eqnarray}}
\newcommand{\be}{\begin{eqnarray}}
\newcommand{\ee}{\end{eqnarray}}

\def\del{\partial}                              

\def\frac#1#2{{#1 \over #2}}
\def\simge{\mathrel{%
   \rlap{\raise 0.511ex \hbox{$>$}}{\lower 0.511ex \hbox{$\sim$}}}}
\def\simle{\mathrel{
   \rlap{\raise 0.511ex \hbox{$<$}}{\lower 0.511ex \hbox{$\sim$}}}}

\def\be{\begin{equation}}
\def\ee{\end{equation}}
\def\bea{\begin{eqnarray}}
\def\eea{\end{eqnarray}}

\begin{document}

\vspace*{3cm}
\title{THE COLOUR GLASS CONDENSATE \footnote{Talk given at 37th Rencontres de Moriond on QCD and Hadronic Interactions, Les Arcs, 16-23 Mar 2002.}
 }

\author{Edmond IANCU}

\address{ Service de Physique Theorique, CE Saclay, F-91191
        Gif-sur-Yvette, France}

\maketitle\abstracts{I briefly review the physical picture of the
saturated gluons at small--$x$ as a  Colour Glass Condensate, and the
effective theory which forms the basis of this picture.}


\section{Introduction: High-energy scattering and the small-$x$ problem}

Understanding the high-energy behaviour of hadronic interactions
from first principles represents a major challenge for theoretical 
particle physics. This problem is intimately
related to that of high parton densities: At high energy, 
the QCD cross-sections are controlled by the ``small--$x$'' gluons
in the hadron wave-function --- i.e., the gluons with small longitudinal
momenta ---, whose density grows rapidly with the
energy (or with decreasing $x$), because of the enhanced radiative processes.
(See Ref. [1] for recent reviews and more references).

Even though QCD is asymptotically free, the high-energy behaviour is
not necessarily perturbative:
It is the transferred momentum  $Q^2$ (and not the
center-of-mass energy squared $s$)
which controls the running of the QCD coupling $\alpha_s(Q^2)$.
In fact, one expects the total cross sections at very large energies
to be controlled by the soft, non-perturbative physics. But perturbation 
theory may still apply to the {\it evolution} of these cross-sections
with $s$.

However, ordinary perturbation theory fails to meet this expectation.
By resumming the dominant radiative corrections at high energy,
the BFKL equation leads to an expression for the gluon density
which grows like a power of $s$, and which, 
with increasing $s$, is driven towards softer
and softer transverse momenta $Q^2$, where perturbation theory cannot
be trusted any longer (``infrared diffusion'').
The power-law increase 
of the gluon distribution with $s$ entails a similar law for the total
cross-section, 
which thus violates the Froissart unitarity bound $\sigma\le \ln^2 s$.

But BFKL, and also DGLAP, are {\it linear} evolution equations, 
which neglect the interactions among the small-$x$ gluons. 
With increasing energy, recombination effects favoured by the
high density should become more and more important, and eventually
lead to a {\it saturation} of the parton densities, i.e., a limitation
of their growth with $s$. In terms of scattering, this 
would correspond to the unitarization of the scattering amplitudes 
at fixed impact parameter.

For given $Q^2$, non-linear effects should become important
when the energy is sufficiently high for the
gluons to overlap by a factor $1/\alpha_s$ (to compensate for the
smallness of their interactions $\propto \alpha_s$). Equivalently,
for a given energy, saturation should occur for those
gluons having a sufficiently large transverse size $1/Q^2$, larger than
the critical value $1/Q^2_s(x)$ at which the ``packing factor''
becomes $\sim 1/\alpha_s$. This requires sufficiently low transverse
momenta $Q^2\simle Q^2_s(x)$, with :
\be\label{Qs}
Q^2_s(x)\,\simeq\,\frac{\alpha_s N_c}{N^2_c-1}\,\frac{
x G(x,Q^2_s(x))}{\pi R^2},\ee
where $x G(x,Q^2)=dN/d\tau$ is the gluon
distribution, i.e.,  the number 
of gluons with longitudinal momentum fraction $x$
and transverse size $\Delta x_\perp \sim 1/Q$ per unit {\it rapidity}
$\tau\equiv \ln(1/x)\sim \ln s$.
Eq.~(\ref{Qs}), together with the BFKL prediction
$x G(x,Q^2) \sim s^{\omega}
$, with $\omega =4\bar\alpha_s\ln 2$ ($\bar\alpha_s\equiv \alpha_s N_c/\pi$),
leads to the conclusion that the {\it saturation scale} $Q^2_s(x)$
should increase as a power of the energy, and also as a power
of the atomic number $A$ (for a nucleus): 
\be\label{QsxA}
Q^2_s(x,A) \approx \Lambda^2_{QCD} 
\,x^{-\lambda}A^\delta,\ee
where $\delta\approx 1/3$ and 
$\lambda \approx 4.84\bar\alpha_s$ 
in the BFKL approximation \cite{AM,SCAL}.
Phenomenological fits of the $F_2$ data at HERA  using saturation
\cite{geometric} lead to a somewhat smaller value $\lambda \approx 0.3$.

Eq.~(\ref{QsxA}) has an important consequence: for
sufficiently large $A$ and/or high enough energy, the saturation scale
is a {\it hard} scale, $ Q^2_s\gg \Lambda^2_{QCD}$, 
so weak coupling methods should be 
applicable\footnote{We assume that the QCD coupling is running with 
$Q^2_s(x)$, which is reasonable since the saturated gluons have 
typical momenta of order $Q_s$.}: $\alpha_s( Q^2_s)\ll 1$. 
This opens the way towards perturbative studies of the high energy limit.

But although the coupling is small, standard perturbative techniques
fail to apply because of the high-density effects, which call
for all-order resummations. A crucial observation, which allowed
for significant technical progress and a deeper physical insight,
is that the small-$x$ regime is a {\it semi-classical} regime, because of the
large occupation numbers \cite{MV94}. This leads one to treat
the small-$x$ gluons as the classical Weizs\"acker--Williams field 
radiated by a random colour charge distribution: that
of the {\it fast} partons with larger $x$. 

With increasing energy, or decreasing $x$, new quantum modes become
relatively ``fast'' and must be included in the colour source seen
by the external probe. Thus, the classical description of 
the small-$x$ gluons is to be seen as an {\it effective theory}
valid at a given value of $x$, and whose ``action'' is evolving
with $x$ \cite{JKLW97,PI,Cargese}. The definitive form of the
equation describing this evolution has been given in Ref. [7]
(see also Ref. [8]),
where the interpretation of the saturated gluons as a {\it Colour Glass
Condensate} has been also proposed. It is my purpose in this talk
to briefly review this physical picture and its mathematical formulation.

\section{The effective theory for the  Colour Glass Condensate}

The classical field equations read:
\beq
(D_{\nu} F^{\nu \mu})_a(x)\, =\, \delta^{\mu +} \rho_a(x^-,x_\perp)\,
\label{cleq}
\eeq
where the colour current in the r.h.s. has just a plus component
since the fast partons are moving at nearly the speed of light
in the positive $z$, or $x^+$, direction.
(I use light-cone vector notations: $x^\pm=(t\pm z)/\sqrt{2}$, 
$x_\perp=(x,y)$, and similarly for the other vectors.) 

The colour charge density
$\rho_a(x^-,x_\perp)$ is localized near the light-cone ($x^-\simeq 0$), 
because of Lorentz contraction, and is {\it frozen}, i.e., it is
independent of the (light-cone) time
$x^+$, because the dynamics of the fast partons is slowed down by
Lorentz time dilation. In other terms, the changes in the configuration
of the fast partons occur over a time scale which is
much larger than the duration
of a collision at small-$x$. This allows for a kind of Born-Oppenheimer
approximation, in which one studies first the dynamics of the classical
fields for a given configuration $\rho_a(x^-,x_\perp)$ of the colour
sources, and then one averages over all the possible configurations.
The weight function $W_\tau[\rho]$ for this averaging
is obtained by integrating out the fast partons,
so it depends upon the rapidity
scale $\tau\equiv \ln(1/x)$ at which one considers the effective theory.

For instance, the unintegrated gluon distribution is obtained as (with
$k^+=xP^+=P^+{\rm e}^{-\tau}$):
\be\label{TPS}
\varphi_\tau(k_\perp)\,\equiv\,
\frac{d^3 N}{d\tau d^2k_\perp}\,=\,\frac{1}{4\pi^3}\,
\langle F^{+i}_a(k^+,k_\perp)F^{+i}_a(-k^+,-k_\perp) \rangle_\tau,\ee
where $F^{+i}_a=\partial^+ A^i_a$ is the ``electric field'' in the
LC-gauge $A^+_a=0$, and the matrix element in the r.h.s. is computed
in the classical theory as (with $\vec x=(x^-,x_\perp)$):
\be\label{FF}
\langle F^{+i}_a(\vec x)F^{+i}_a(\vec y)\rangle_\tau\,=\,
\int { D}[\rho]\,\,W_\tau[\rho]\,
{\cal F}^{+i}_a(\vec x)
{\cal F}^{+i}_a(\vec y)\,,\ee
where ${\cal F}^{+i}_a\equiv {\cal F}^{+i}_a[\rho]$ is the 
solution to the classical Yang-Mills equations (\ref{cleq}) in this LC gauge.
At small $x$, the colour fields become strong,
$F^{+i}_a\sim 1/g$, so the classical solution must be computed exactly.
Given the special geometry of the source, 
this exact solution is known indeed \cite{Cargese} :
\be\label{F}
{\cal F}^{+i}(\vec x) \simeq \delta(x^-)\frac{i}{g}\,V(\del^i V^\dagger)
(x_\perp),\qquad V^\dagger(x_{\perp})\equiv {\rm P} \exp
 \left \{ig \int dz^-\,\alpha_a (z^-,x_{\perp}) T^a
 \right \},\ee
with $\alpha^a(\vec x)$ the solution to the 2-dimensional Poisson equation
$- \nabla^2_\perp \alpha_a({\vec x})=\rho_a({\vec x})$
(this is the ``Coulomb'' field in the infinite momentum frame).
In terms of the ordinary electric ($E^i_a$) and magnetic ($B^i_a$)
fields, eq.~(\ref{F}) describes a plane-wave--like
configuration in which ${\bf E}$ and ${\bf B}$
are transverse to each other and also to the direction of propagation,
and have equal magnitudes.

The average over $\rho$ in 
eq.~(\ref{FF}) is reminiscent of that performed for systems
with a frozen disorder, like {\it spin glasses}. A spin glass is a collection
of magnetic impurities (the ``spins'' $S_i$) randomly distributed in some
non-magnetic host, with lattice points $i,\,j,..$. 
The disorder can be characterized by
treating the spin-spin couplings --- the ``link variables''
$J_{ij}$ --- as fixed, but random. In reality, the 
$J_{ij}$'s can change with time, but their changes occur only on time scales
much larger than any characteristic time scale for the dynamics of the spins.
Thus, when studying, e.g., the thermalization of the spin system,
one can assume that thermal equilibrium is reached for each given 
configuration of the $J_{ij}$'s, and then average over the latter.
The free energy is obtained as:
\be \label{SG-F}
F\,=\,-T \int D[J]\,W[J]\,\ln Z[J]\,,\qquad 
Z[J]\,=\,\sum_{\{S\}}\,{\rm e}^{-\beta H_J[S]},\ee
where $H_J[S]=-\sum_{i,j} J_{ij} S_i S_j$ and $W[J]$ is the
weight function for the link variables.
Clearly, there is a formal analogy between eqs.~(\ref{SG-F}) and (\ref{FF}),
with $J_{ij}\leftrightarrow \rho_a(\vec x)$ and 
$S_i\leftrightarrow F^{+i}_a(\vec x)$.
In this analogy, spin is replaced by colour, so one can characterize the
small-$x$ gluonic matter described by the effective theory in
eqs.~(\ref{cleq}) and (\ref{FF}) as a ``colour glass'' \cite{PI}.

It is also instructive to contrast this physical situation 
to what happens in a {\it plasma}. There, the charged
particles are very mobile, so they can rapidly adapt themselves
to the changes in the electromagnetic background field $A^\mu$.
It is then appropriate to first compute the induced current
$j^\mu[A]$ by ``integrating out'' the charged particles in the background
of a given field $A^\mu$, and only then solve the Maxwell equations 
with source $j^\mu[A]$. That is, for a plasma,
the analogs of the operations
in eqs.~(\ref{cleq}) and (\ref{FF}) above are performed in the
{\it opposite} order.

\section{Non-linear evolution and saturation}

When the rapidity $\tau$ is increased by $d\tau$, i.e., the hadron
is further accelerated, the quantum gluons with rapidities $\tau'$
in the interval $\tau<\tau'<\tau+d\tau$ get effectively frozen
(since they are time dilated w.r.t. a collision at the new rapidity scale
$\tau+d\tau$), and therefore must be incorporated in the effective theory.
They become a part of the colour glass.

This entails a change in the properties of the colour source $\rho$
--- its support and its correlation functions $<\rho(1)\rho(2)\cdots>$
--- which can be absorbed into an appropriate ``renormalization''
of the weight function $W_\tau \to W_{\tau+d\tau}$. The result of
a lengthy analysis in which quantum modes are integrated out in
the background of the colour fields 
${\cal F}^{+i}_a(\vec x)$ generated at the previous steps in the evolution,
is the following, non-linear evolution equation for
$W_\tau[\rho]$ \cite{JKLW97,PI} :
\be\label{RGE}
{\del W_\tau[\rho] \over {\del \tau}}\,=\,
 {1\over 2} \int_{x_\perp,y_\perp}\,{\delta \over {\delta
\rho_\tau^a(x_\perp)}}\,\chi^{ab}(x_\perp, y_\perp)[\rho]\, 
{\delta \over \delta \rho_\tau^b(y_\perp)}\,W_\tau[\rho].\ee
This is a functional Fokker-Planck equation. It describes quantum evolution
towards small $x$ as a random walk on the functional space spanned by
$\rho_a(x^-,x_\perp)$. The kernel 
$\chi^{ab}(x_\perp, y_\perp)[\rho]$, which plays the role of a
``diffusion coefficient'', is the correction
to the 2-point function of $\rho$ induced by the quantum fluctuations.
It is itself a non-linear functional of the original source $\rho$,
upon which it depends via the Wilson lines
$V$ and $V^\dagger$, eq.~(\ref{F}).

The functional equation above can be converted into ordinary evolution
equations for the correlation functions of the Wilson lines.
This generates the same equations as obtained by Balitsky, Kovchegov
and Weigert within different approaches \cite{BKW}. However, progress
can be made also via direct investigations of the functional equation
(\ref{RGE}), for which approximate solutions have been obtained \cite{SAT}.
These are briefly described now:

 When the source is weak,
corresponding to a low-density system, or high transverse momenta
$ k_\perp^2 \gg Q_s^2(\tau)$, the Wilson lines can be expanded
to lowest order in the Coulomb field $\alpha_a$ 
(e.g., $V^\dagger(x_{\perp})\approx 1 + ig \int dx^-
\alpha_a (x^-,x_{\perp}) T^a$), 
and eq.~(\ref{RGE}) reduces to the linear BFKL equation for
the unintegrated gluon distribution of eq.~(\ref{TPS}).
This leads to the usual, exponential, increase with the rapidity:
$\varphi_\tau(k_\perp)\propto {\rm e}^{\omega\tau}$.

However, when the density becomes as large as 
$\varphi_\tau(k_\perp)\sim 1/\alpha_s$ --- this happens at rapidities
$\tau$ such as $k_\perp^2\simle Q_s^2(\tau)$ 
(cf. eqs.~(\ref{Qs})--(\ref{QsxA})) ---, then the colour fields
are strong $\alpha_a\sim 1/g$, so the Wilson are rapidly oscillating and
average away to zero: $V\approx V^\dagger \approx 0$. Then the
r.h.s. of eq.~(\ref{RGE}) simplifies drastically: The kernel $\chi$
becomes independent of $\rho$ : $\chi(k_\perp)=(1/\pi) k_\perp^2$, and
vanishes as $k_\perp^2\to 0$, which shows that {\it colour neutrality}
is achieved  over an area $\simge
1/Q_s^2(\tau)$.

With this simplified kernel, eq.~(\ref{RGE}) is readily solved,
and the corresponding gluon distribution (\ref{TPS}) can be then computed.
The final result reads \cite{SAT} (up to a numerical factor)  :
\be\label{SAT}
\varphi_\tau(k_\perp)\,\simeq\,\frac{N^2_c-1}{N_c}\,{1\over\alpha_s}\,
\ln{Q_s^2(\tau)\over k_\perp^2}\,\propto \,\tau,\qquad
{\rm for}\,\,\,k_\perp\ll Q_s(\tau),\ee
and exhibits gluon {\it saturation} : At low momenta
$k_\perp\simle Q_s(\tau)$, the
gluon density grows only linearly with $\tau$, i.e.,
{\it logarithmically} with the energy. Thus,
with increasing $s$, the new partons are predominantly produced at large
transverse momenta $\simge  Q_s(\tau)$.

This suggest that saturation is a natural mechanism 
to restore {\it unitarity} :
For an external probe with given 
resolution $Q^2$, the scattering amplitude $T$
saturates, $T=1$, at any impact parameter where the condition
$Q^2< Q_s^2(\tau)$ is satisfied. But to convincingly
demonstrate the unitarization of the total cross-section, one
must still show that the area of the ``black disk'' where 
$T=1$ is not increasing faster than $\ln^2s$. 
This requires extending the previous analyses
of eq.~(\ref{RGE}) by including the dependence upon the impact parameter.

\end{document}